# A nanophotonic all-optical diode for non-reciprocal transmission of circularly polarized lights

Hongming Fei,[1,2,*,†] Min Wu,[1,2,†] Han Lin,[3,*] Xin Liu,[1,2] Yibiao Yang,[1,2] Mingda Zhang,[1,2] and Binzhao Cao[1,2]

[1] School of Physics and Optoelectronics, Taiyuan University of Technology, Taiyuan 030024, China
[2] Key Laboratory of Advanced Transducers and Intelligent Control System, Ministry of Education, Taiyuan University of Technology, Taiyuan 030024, China
[3] Micro-Photon Center, Swinburne University of Technology, 3122, Australia

**Abstract** All optical diodes (AODs) play an important role in quantum optics and information processing, in which the information is encoded by photons. Only circularly polarized lights are able to carry the spin states of photons, which has been intensively used in quantum computing and information processing and enable new research fields, such as chiral quantum optics. An ideal AOD should be able to work with arbitrary polarizations states, including circularly polarized lights, which has not been demonstrated yet. In this paper, we theoretically demonstrate for the first time a nanophotonic AOD that is able to work with circularly polarized lights. The AOD nanostructure is based on a heterostructure of two-dimension silica and silicon photonic crystals (PhCs). By controlling the effective refractive indices of the PhCs and using an inclined interface, we are able to exploit generalized total reflection principle to achieve non-reciprocal transmission of circularly polarized lights. In addition, the nanophotonic AOD is able to achieve high forward transmittance (>0.6) and high contrast ratio (~1) in a broad wavelength range (1497 nm～1666 nm). The designed nanophotonic AOD will find broad applications in optical quantum information processing and computing.

[*] Corresponding authors: feihongming@tyut.edu.cn；hanlin@swin.edu.au
[†] These authors contributed equally to this work.

## I. INTRODUCTION

Advanced photonic nanostructures are currently revolutionizing the optics and photonics that underpin applications ranging from light technology to quantum-information processing. All optical diodes (AOD) based on photonic nanostructures [1] has recently attracted broad attentions, as they play a key role in quantum-information processing, as well as the possibility in achieving integrated photonic circuits. The polarization states of light is of crucial importance in quantum optics and information processing [2]. Among different polarizations states, circular polarizations are utilized in various applications as they carry the spin of photons, including chiral optics [3, 4] and spin-state control in quantum information technology [5-7]. The recent advancement in chiral quantum optics based on circularly polarized lights offers fundamentally new functionalities and applications, such as non-reciprocal single-photon coupling and complex quantum networks. Therefore, a nanophotonic AOD that is able to achieve non-reciprocal transmission of circularly polarized photons is an essential element in chiral quantum optics.

It has been demonstrated different designs of AODs [8-15] based on nanostructures. However, none of the proposed designs are able to achieve non-reciprocal transmission of circularly polarized lights due to the polarization dependent transmission properties. Therefore, it remains a great challenge in achieving nanophotonic AOD for circularly polarized lights. Herein, we present a nanophotonic AOD for circularly polarized lights based on a two-dimensional (2D) photonic crystal (PhC) heterostructure, which consists of two square-lattice PhCs ($PhC_1$ and $PhC_2$). The effective refractive indices of the PhCs are controlled by the materials and the structure parameters. In addition, an inclined interface is applied to exploit generalized total reflection principle to achieve non-reciprocal transmission of circularly polarized lights. We further optimize the structure parameters to improve the performance of the AOD device. As a result, the nanophotonic AOD is able to achieve high forward transmittance (>0.6) and high contrast ratio (~1) in a broad wavelength range (1497 nm~1666 nm). The designed nanophotonic AOD achieves non-reciprocal transmission of circularly polarized light and will find broad applications in chiral quantum optics.

## II. THEORETICAL MODEL AND PHYSICAL MECHANISM ANALYSIS

The schematic of the structure is shown in Fig. 1, in which a heterostructure consists two square-lattice PhCs ($PhC_1$ and $PhC_2$). Two PhCs have the same lattice constant $a$, but different air-hole shapes (circular and square, respectively). A circularly polarized light is considered as the incident light of the AOD. The goal is to achieve high forward (along the positive $x$ direction) transmittance of the circularly polarized light over a broad wavelength range, meanwhile completely block the backward (along the negative $x$ direction) transmittance. The performance of the AOD is characterized by the forward transmittance ($T_F$) and contrast ratio ($C$), which is defined by

$$C = (T_F - T_B)/(T_F + T_B) \qquad (1)$$

where $T_B$ is the backward transmittance.

For the circularly polarized lights, the electric field vector rotates at a constant angular velocity $\omega$ (the circular frequency of the light wave) regarding the optical axis, which is a combination of two linearly polarized lights with $\pi/4$ phase difference. The electric field vectors of circularly polarized lights are propagating along the $x$ axis can be written as

$$E = A[\cos(\omega t - kx)e_y \mp \sin(\omega t - kx)e_z] \quad (2)$$

when the + and – correspond to left (LCP) and right handed circular polarized (RCP) lights, respectively. From the Eq. (2) one can see that there are four requirements to achieve forward transmission of circularly polarized light :

1) The forward transmittance ($T_F$) of the two linearly polarized lights, including transverse electric (TE) and transverse magnetic (TM) modes, should be equal.
2) The polarization states of the transmitted TE and TM polarization should be maintained.
3) The phase difference ($\pi/4$) of the TE and TM polarizations should be the maintained.
4) The intensity distribution of the TE and TM polarizations should be spatially overlapped.

In this way, the recombination of the two transmitted linearly polarized lights can be circularly polarized.

In addition, it is required to simultaneous completely block the TE and TM polarizations to achieve minimized backward transmittance ($T_B$) and high contrast ratio ($C$).

Currently, there is no demonstration of polarization independent non-reciprocal transmission in any nanophotonic AOD structure to the best of our knowledge, as nanophotonic structures usually work for only one polarization according the required coupling conditions. Therefore, it is extremely challenge to meet all four requirements at the same time. We design the AOD consists of two 2D PhCs (PhC$_1$ and PhC$_2$), which are air holes embedded in silica (PhC$_1$) and silicon (PhC$_2$) materials, respectively. The radius of circular air-hole is $0.29a$, and the half-side length of square air-hole is $0.23a$. The refractive indices of silica and silicon are 1.495 and 3.48 at the telecommunication wavelength (1550 nm). An inclined interface is placed between the two PhCs to apply the total internal reflection principle, which along the Γ-M of the PhCs. According to the above mentioned requirements, we first optimize the transmission properties of TE/TM polarized lights, including forward transmittance, phase modulation, intensity distribution and contrast ratio, respectively. Then we study the non-reciprocal transmission properties of circularly polarized lights. As the requirements for LCP and RCP are exactly the same, we take RCP as example.

As shown in Fig. 1, the light couples into the PhC$_1$ along the Γ-X direction ($x$-direction) and exit along the Γ-M. Then the light enters the PhC$_2$ along the Γ-M direction and couples out along the Γ-X direction. Therefore, it requires the PhC$_1$ and PhC$_2$ has high transmittance for both TE and TM polarizations along both Γ-X and Γ-M direction at the target wavelength 1550 nm (corresponding to normalized frequency of 0.316 a/λ). To investigate light transmission property, the TE/TM band structures of PhC$_1$ (circular hole) and PhC$_2$ (square hole) are calculated by the plane wave expansion (PWE) method. The results are shown in Fig. 2. The blue lines represent the TE bands, and the red lines represent the TM bands. As seen from Fig. 2(a) and 2(b), PhC$_1$ is pass band along the Γ–X direction and PhC$_2$ has stop band in TM mode along Γ–X directional in the frequency of 0.316 a/λ marked by green horizontal lines.

To further study the light propagation in PhC$_1$ and PhC$_2$, we plot the equal frequency contours (EFCs) for both polarizations of the two PhCs, as the propagation direction of light waves in the PhCs is along the gradient direction of the equal frequency lines. The EFCs are centered on the Γ point, the black arrow (red arrow) in Fig. 2(c) to 2(f) represents the forward incidence (the backward incidence) and the direction of light propagation in PhCs. The light can go straight into PhC$_1$ along Γ–X direction and reaches the interface can be seen in Fig. 2(c) and 2(e). The wave vectors of the incident light refract at the interface because of different effective refractive indices of the PhCs, then is propagating along the Γ–X direction in TE mode and the Γ–M direction in TM mode in PhC$_2$, indicating by the black arrow in Fig. 2(d) and 2(f). While the backward transmission in the frequency of 0.316 a/λ in TE mode, the light wave in Γ–X direction for PhC$_2$ is propagating along the 90° direction in wide band as shown in Fig. 2(d), thus the light is completely cut off. For TM polarization there is stop band along the Γ-X direction in PhC$_2$, thus the backward propagating light is forbidden. The analysis confirms that the heterostructure has non-reciprocal transmission characteristic in theory.

### III. TRANSMISSION CHARACTERISTIC IN 2D/3D HETEROSTRUCTURE

In addition, it is necessary to maintain the original polarization states of the TE and TM polarizations to effectively transmit the circularly polarized light. To visualize the non-reciprocal transmission, the TE/TM field intensity distributions of the forward and backward propagating lights are simulated by the use of the finite-difference time-domain (FDTD) method at 0.316 a/λ in Figs. 3. From Figs. 3(a) and 3(b), one can see that both the TE and the TM polarization are able propagate along the Γ-X direction with high efficiency. The light from the right side can't propagate into PhC$_1$ because of the above-mentioned mechanisms. Through broadening the width of the right side waveguide, fully collecting the transmission light wave of a specific wavelength range, it can increase the forward transmission and improve the unidirectional transmission performance of the hererostructure. Meanwhile, adjusting the distance of two sides photonic crystals from the interface so that the efficient unidirectional transmission of heterostructure are realized. From optimization we found that optimal width of waveguide in the right side is $D = 9 \mu$m in FIG. S1 and the distance of two sides photonic crystals from the interface $a$ and 1.5 $a$ in FIG. S2, non-reciprocal transmission is the most effective. It can be seen from Fig. 3(c), in TE polarized mode, we can optimize the structure to achieve high forward transmittance (>0.59) and high contrast ratio (>0.97) in the wavelength range of 1502 nm～1800 nm. Forward transmittance is 0.67 and contrast ratio is 0.99 in the wavelength of 1550 nm. In the TM polarized mode, the high forward transmittance (>0.6) and the high contrast ratio (>0.97) in the wavelength range of 1472 nm～1668 nm, Forward transmission is 0.64 and contrast ratio is 0.99 in the wavelength of 1550 nm can be shown in Fig. 3(d).

In order to visually observe the non-reciprocal transmission performance of 3D photonic crystal heterostructure device (Fig. 4(d) inset), the electric field intensity of propagating light are simulated by FDTD method at 0.316 a/λ. It can be seen from Fig. 4(a), the forward transmitted light from the left side can highly efficient propagate along the Γ–X direction in $PhC_1$, while the backward light is inhibited in the $PhC_2$. Owing to the optical refraction and photonic crystal transmission mechanism in $PhC_2$ for the designed frequencies, forward transmission is realized. And the light from the backward can't propagate into $PhC_1$ because of total reflection. In the Fig. 4(b), we simulated field distribution and light polarization state of *y-z* cross-section at the output port, different locations inside the PhCs support different superpositions and phase of E*y* and E*z*. At each point the field may be expressed as polarization circles or ellipses. We can see the polarization of light varies slightly from circular to slightly elliptical. The lights maintain circular polarization around the center of the output port where the peak intensity locates.

We simulates transmittance of 2D/3D heterostructure for circularly polarized light. The results are shown in Figs. 4(c) and 4(d). We optimize the 2D heterostructure to achieve high forward transmittance (>0.6) and high contrast ratio (>0.97) in the wavelength range of 1497 nm～1666 nm in a broad bandwidth (up to 169 nm). Forward transmission is 0.65 and contrast ratio is 0.99 in the wavelength of 1550 nm for circularly polarized incident light. An ideal 2D photonic crystal is infinitely thick along the *z*-direction, however, the realistic structures are with finite thickness in the vertical *z*-direction [16-18], known as photonic-crystal slabs. We design a 2D photonic crystal device with a thickness of 1500 nm. In the 3D heterostructure, the high forward transmittance (>0.55) and the high contrast ratio (>0.98) in the wavelength range of 1419 nm～1632 nm (213 nm) for circularly polarization light is achieved. Forward transmittance and contrast ratio of the 3D heterostructure is 0.57 and 0.99 in the wavelength of 1550 nm, which shows high performance. Therefore, the efficient non-reciprocal transmission for circularly polarization light beam in heterostructure is achieved.

## IV. CONCLUSION

In conclusion, we have demonstrated a nanophotonic AOD for circularly polarized lights based on a two-dimensional (2D) PhC heterostructure, which consists of two square-lattice PhCs. The transmission mechanism analysis and optimization of heterostructure are studied, and the resulted the AOD achieves forward transmittance (>0.6) and high contrast ratio (~1) in the wavelength range of 1497 nm～1666 nm (169 nm). The demonstrated nanophotonic AOD opens up new possibility in designing new chiral quantum optics devices for quantum computing and information processing, as well as integrated optics and all optical network.

## ACKNOWLEDGMENTS

The authors would like to thank the financial supports from National Natural Science Foundation of China (Grant No. 61575138), the Young Scientists Fund of the National Natural Science Foundation of China (Grant No. 61505135), and the Natural Science Foundation of Shanxi Province, China (Grant No. 2016011048).

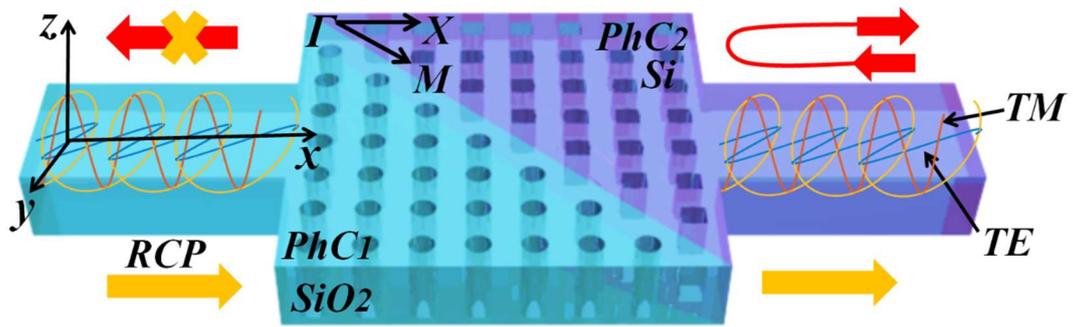

FIG. 1. The sketch map of right-handed circularly polarized light propagates along *x* axis and photonic crystal heterostructure with air holes, circularly polarized light composed by transverse electric (TE) and transverse magnetic (TM) polarized lights.

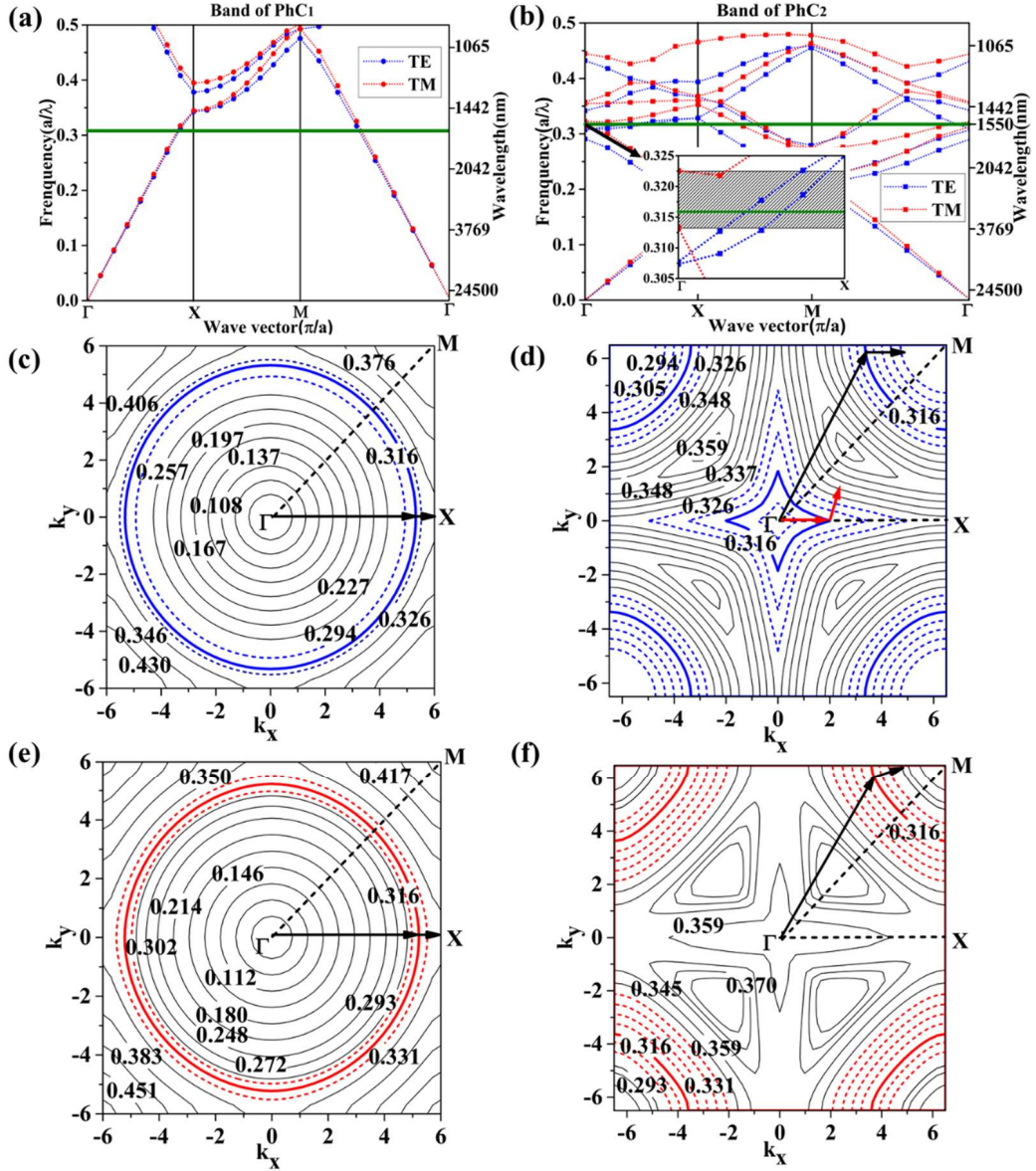

FIG. 2. Band diagrams of PhC₁(a) and PhC₂(b); (c) the EFCs of PhC₁ in the first band (blue lines represent 0.316 a/λ, blue dashed lines represent one way transmission band in TE mode); (d) the EFCs of PhC₂ in the fourth band; (e) the EFCs of PhC₁ in the first band (red lines represent 0.316 a/λ, red dashed lines represent one way transmission band in TM mode); (f) the EFCs of PhC₂ in the fourth band.

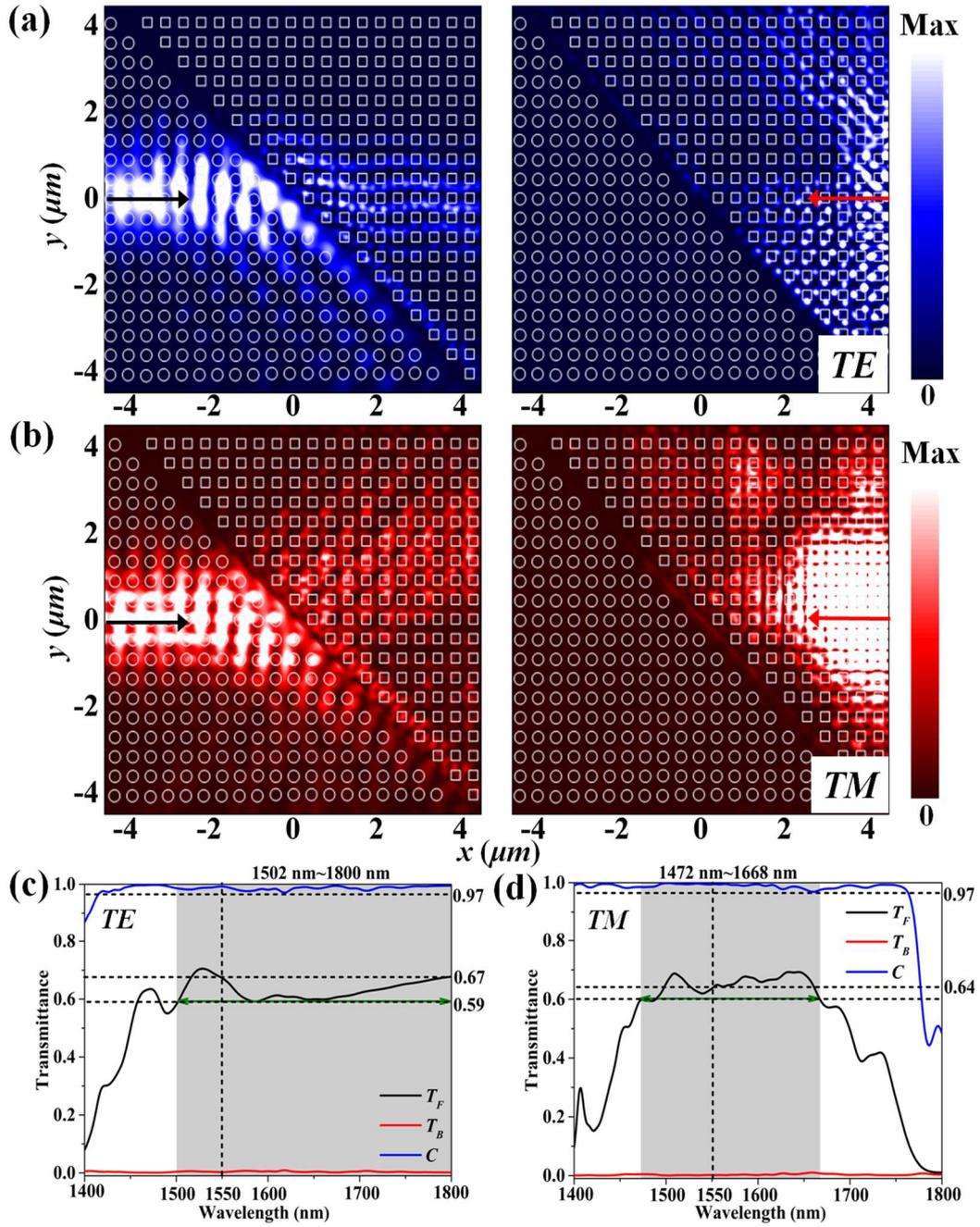

FIG. 3. In 2D photonic-crystal heterostructure, the electric field intensity of opposite directions at 0.316 a/λ (a) forward and backward in TE polarized mode; (b) forward and backward in TM polarized mode. Transmittance and contrast ratio of 2D heterostructure in TE/TM polarized mode (c) and (d).

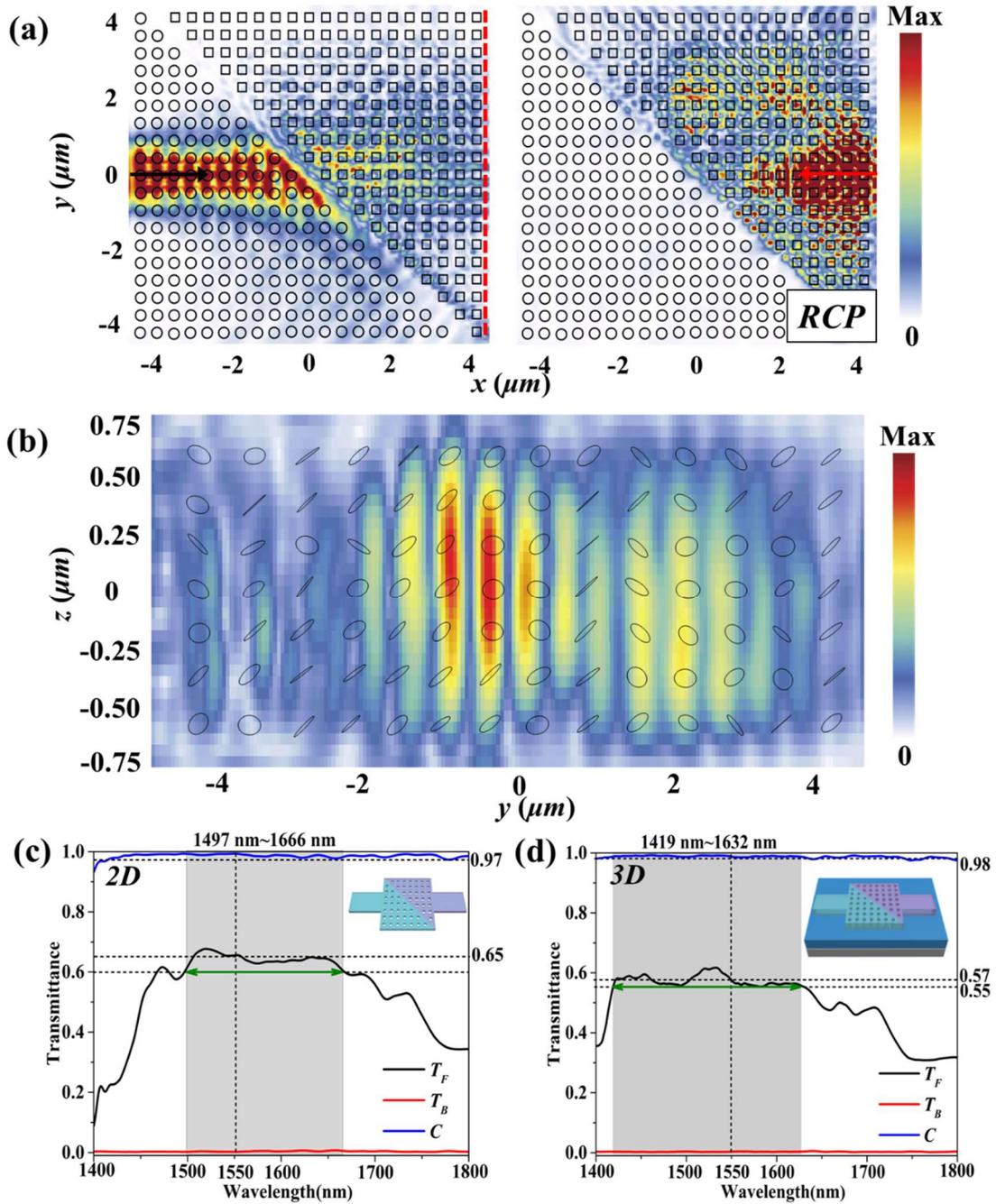

FIG. 4. RCP light propagates in 3D heterostructure, (a) the electric field intensity of opposite directions at 0.316 a/λ (red dashed line represents cross-section of the device); (b) Electric field distribution of *y-z* cross-section at the output port, red marks show polarization circle and polarization ellipse. Transmittance and contrast ratio of 2D/3D heterostructures (c) 2D structure; (d) 3D structure.

# Supplementary materials for A nanophotonic all-optical diode for non-reciprocal transmission of circularly polarized lights

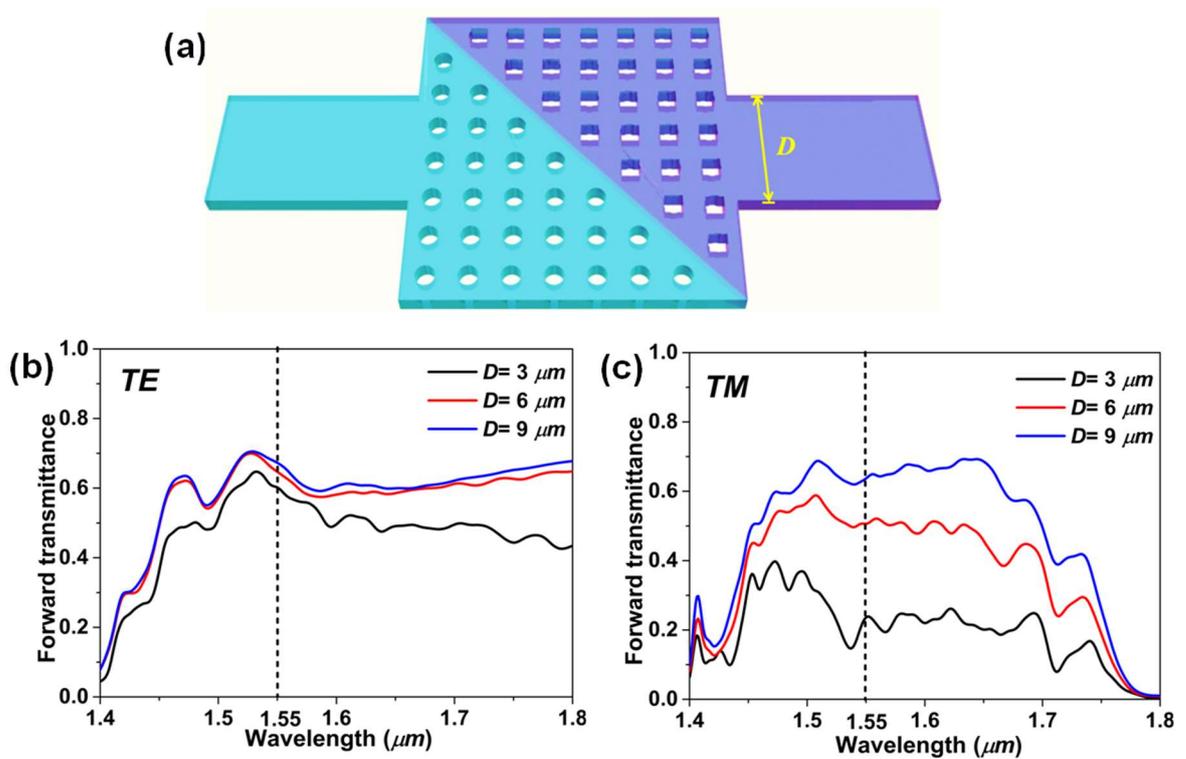

FIG. S1 The sketch map of photonic crystal heterostructure with air holes (a). When broadening the width of the right side waveguide $D$, forward transmittance of the heterostructure with different $D$ in TE/TM polarized modes (b) and (c).

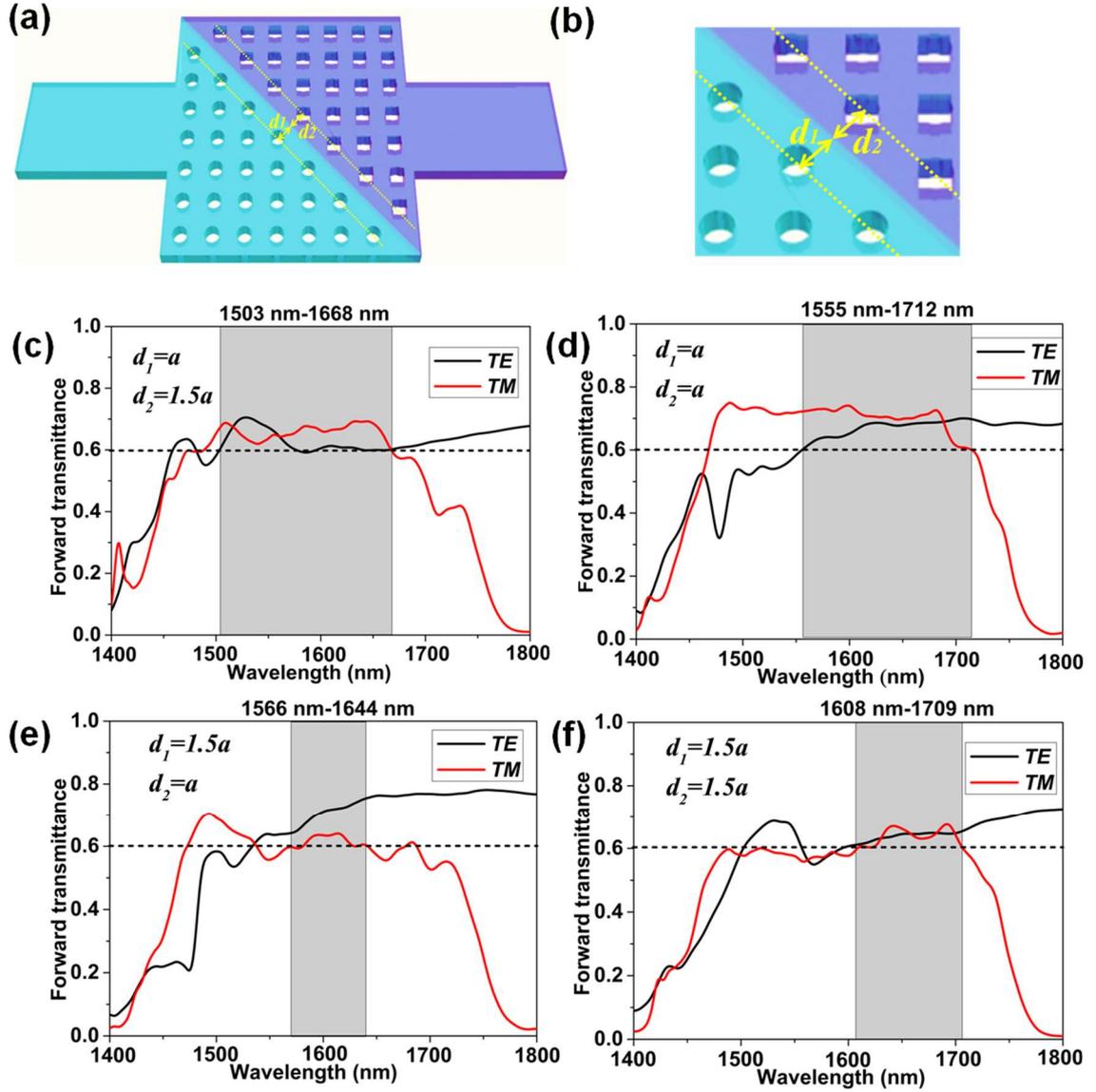

FIG. S2 The sketch map of photonic crystal heterostructure with air holes and the partial enlarged detail of heterointerface (a) and (b). Forward transmission spectra for the TE/TM mode with different level distances between interface and adjacent air holes in the heterostructure: (c) $d_1 = a$, $d_2 = 1.5a$; (d) $d_1 = a$, $d_2 = a$; (e) $d_1 = 1.5a$, $d_2 = a$; (f) $d_1 = 1.5a$, $d_2 = 1.5a$ (grey region represents one transmission band, which is forward transmission is higher than 0.6).